# Transport of magnetically sensitive atoms in a magnetic environment

Davlet Kumpilov[1,2], Ivan Pyrkh[1,2], Ivan Cojocaru[1,3], Polina Trofimova[1], Arjuna Rudnev[1,2], Vladimir Khlebnikov[1], Pavel Aksentsev[1,4], Ayrat Ibrahimov[1,2], Alexander Yeremeyev[1,5], Kirill Frolov[1,2], Sergey Kuzmin[1,2], Anna Zykova[1], Daniil Pershin [1,3], Vladislav Tsyganok [1], Alexey Akimov[1,3]

[1]*Russian Quantum Center, Bolshoy Boulevard 30, building 1, Skolkovo, 121205, Russia*

[2]*Moscow Institute of Physics and Technology, Institutskii pereulok 9, Dolgoprudny, Moscow Region 141701, Russia*

[3]*PN Lebedev Institute RAS, Leninsky Prospekt 53, Moscow, 119991, Russia*

[4]*Bauman Moscow State Technical University, 2-nd Baumanskaya, 5, Moscow, 105005, Russia*

[5]*Skolkovo Institute of Science and Technology, Bolshoy Boulevard 30, building 1, Skolkovo, 121205, Russia*

email: a.akimov@rqc.ru

Among interesting applications of cold atoms, quantum simulations attract a lot of attention. In this context, rare-earth ultracold atoms are particularly appealing for such simulators due to their numerous Fano-Feshbach resonances and magnetic dipole moments in the ground state. Creating a quantum gas microscope requires a large optical access that may be achieved using transport of atoms between separate vacuum volumes. We demonstrate that in case of the transport of magnetic atoms the magnetic field can be directly measured and adjusted to reduce additional losses after the transport therefore increasing the efficiency of subsequent evaporation cooling. This approach allows to transfer over 85% of the atoms from the main chamber to the scientific chamber, located 38 cm away with moderate laser power of 26 W without atomic polarization decay.

Rare-earth ultracold atoms are emerging as an intriguing platform for quantum simulations[1–8]. The presence of *f*-electrons causes a large magnetic moment in the ground state facilitating long-range interactions[9,10], and a dense spectrum of low-field Feshbach resonances, providing detailed control over short-range interactions[11–15]. Moreover, rare-earth elements allowed to observe quantum droplets[16–19]. Along with cavity-mediated interactions[20] and spin-orbit coupled quantum gases[21], dipolar gases play a noticeable role in the research of supersolids[22–24]. The quantum gas microscope has been recently demonstrated with Erbium atoms[25].

A quantum gas microscope with ultracold atoms requires a high numerical aperture. Although it can be achieved in a single-chamber experiment design[25,26], creation of separate vacuum volumes is more common[27–33]. A number of methods to transfer atoms between two vacuum chambers have been developed. One can use magnetic fields to transport atoms captured in magneto-optical traps (MOTs)[34,35]. Also, shifting the focus of the optical dipole trap (ODT) beam using translation stages[36], focus-tunable lenses[37,38] and 1D optical lattices[39] allows to move atomic clouds. By transporting atoms in the ODT to the volume of the glass cell, a quantum microscope with non-magnetic ytterbium atoms was achieved[40].

Contrarily, when the atoms have large magnetic moment, their transport becomes more complicated due to a number of reasons. First, the presence of an external magnetic field gradient causes an extra force, which affects the motion of the atoms. Next, the high density of Fano-Feshbach resonances typical of rare-earth atoms can cause losses of atoms when the magnetic field changes along the transport trajectory[11,12,15,41]. Finally, non-adiabatic changes in the magnetic field can cause depolarization of an atomic ensemble and subsequent spin-relaxation losses. Although some works reported the optical transport of magnetic atoms with high efficiency, there is no information about these obstacles[13,42]. Perhaps, there were no extra losses during transport because of high power of the ODT beam suppressing the magnetic field forces.

This paper presents a method for reasonably fast optical transport of magnetic dipolar thulium 169 atoms. This method exploits the maintenance of the magnetic field during the transport to preserve the polarization of the atoms at the lowest magnetic sublevel $|F=4, m_F=-4\rangle$ in the ground state $4f^{13}(^2F^0)6s^2$ with corresponding total electronic momentum $J=\frac{7}{2}$ and nuclear spin $I=\frac{1}{2}$. Moreover, sophisticated tailoring of the focus-tunable waist position compensated for the force from the gradient of the magnetic field. This approach allows to avoid additional losses during the transport using less power in the ODT beam. Moreover, the approach reduces the spin-relaxation losses after transport that can be very useful for more efficient evaporative cooling and production of quantum degenerate states in the separate vacuum volume.

The precooling and trapping stages of the experiment closely followed procedures presented in previous studies[43–48]. The pre-cooling of the atoms was realized via the Zeeman slower and 2D optical molasses operated at the strong transition $4f^{13}\left(^{2}F^{0}\right)6s^{2} \rightarrow 4f^{12}\left(^{3}H_{5}\right)5d_{3/2}6s^{2}$ with a wavelength of 410.6 nm and a natural width of $\Gamma = 2\pi\gamma = 2\pi \cdot 10.5\,\text{MHz}$. Following the precooling stage, atoms were loaded into the MOT operating at the weaker transition $4f^{13}\left(^{2}F^{o}\right)6s^{2} \rightarrow 4f^{12}\left(^{3}H_{6}\right)5d_{5/2}6s^{2}$ with a wavelength of 530.7 nm and a natural width of $\Gamma = 2\pi\gamma = 2\pi \cdot 345.5\,\text{kHz}$. Then the reduction of MOT light intensity provided the polarization of atoms at the lowest magnetic sublevel $|F = 4; m_{F} = -4\rangle$ of the ground state[49–53]. The atoms were cooled down to 22.5 ± 2.5 µK and then loaded into the ODT formed by a linearly polarized laser beam waist with a wavelength of 1064 nm.

The MOT was formed in the Kimball Physics MCF800-ExtOct-G2C8A16 UHV vacuum chamber, the scientific chamber was the glass cell. The distance between the centers of the two chambers was measured at 38 cm. To implement the optical transport, the beam waist with the total power of 26 W was created by a system of a current driven focus-tunable lens Optotune EL-16-40-TC-NIR-20D-1-C and a bi-convex lens with focal length $F = 250$ mm (Figure 1a). Variations of the optical power $1/f$ of the focus-tunable lens made it possible to alter the ODT beam waist position by more than 40 cm (Figure 1b) exceeding the distance between the main chamber and the science chamber. The camera mounted on a moving table outside vacuum chamber allowed to measure the ODT beam waist in any position within the transport range. The experimental adjustment of $\delta$ (see Figure 1a) enabled to make the waist constant at the value of 37 µm along the specified range (see Figure 1c) corresponding to the trap depth of 430 µK.

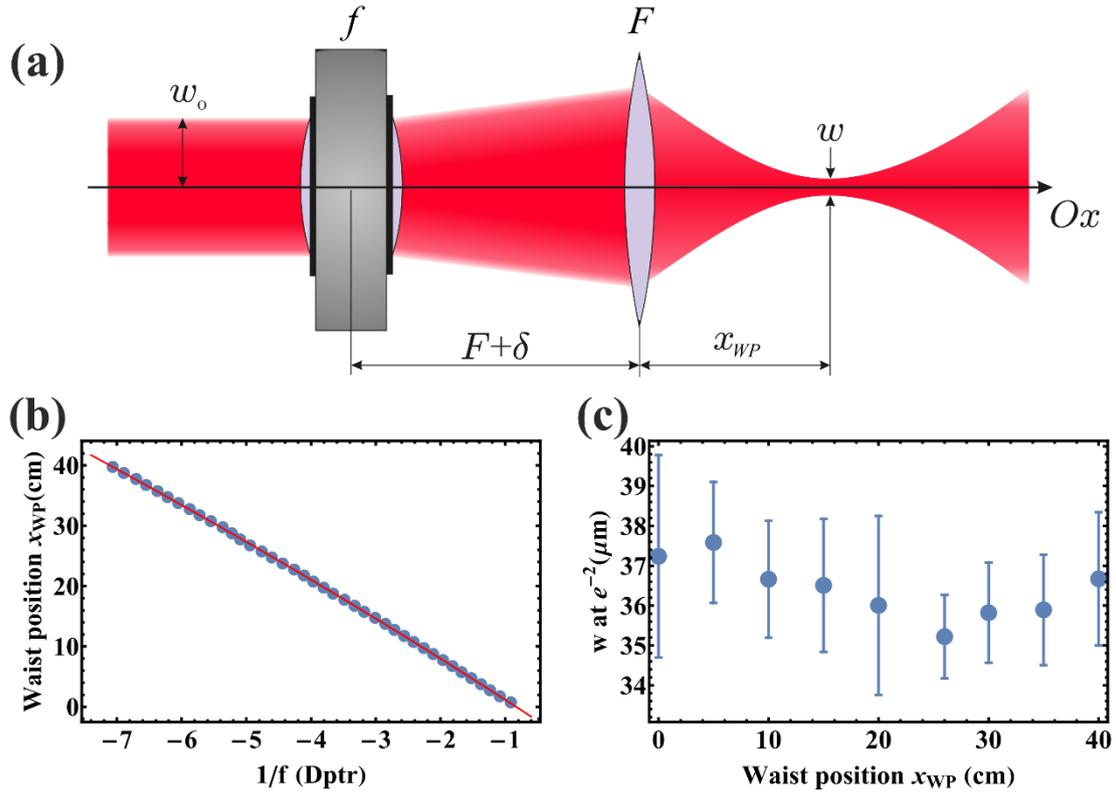

Figure 1. a) Optical scheme of the transport ODT. Initial beam waist $w_0$ is equal to 2.08 mm. Variating $1/f$ of the tunable lens allowed to move the position of the waist $x_{WP}(t)$. Experimentally varying $\delta$ allowed to make the waist value equal to 37 μm in all the waist coordinates. b) Waist coordinate relative to the starting point of the transport versus the optical power of the tunable lens. The solid line is a fit of the experimental data with the result of gaussian beam calculations. The error bars are smaller than the point sizes and are not shown. c) The waist of the ODT versus the waist coordinate.

Shifting the focus of the ODT beam forces the atoms to move along the beam axis. However, all the issues listed in the introduction additionally demand maintaining the magnetic field during the motion. There are at least two limitations on the magnetic field value. On the one hand, previous results revealed a remarkable depolarization in a low magnetic field[54], possibly because the anisotropy of dipole-dipole interactions lead to spin non-conserving collisions [55]. To evaluate a threshold magnetic field sufficiently suppressing the depolarization collisions, the average polarization $\langle m_F \rangle$ was assumed to obey the Boltzmann formula:

$$\langle m_F \rangle = \frac{1}{Z(B)} \sum_{m_F=-4}^{m_F=4} m_F Exp\left[-\frac{m_F \mu_B gB}{k_B T}\right], \quad Z(B) = \sum_{m_F=-4}^{m_F=4} Exp\left[-\frac{m_F \mu_B gB}{k_B T}\right], \quad (1)$$

where $B$ is the magnetic field absolute value, $T$ is the temperature of the atomic cloud, and $g$ is the Lande factor. For $T < 50\,\mu K$ and $\langle m_F \rangle < -3.95$, the value of $B$ should exceed 2.27 G (Figure 2). On the other hand, the presence of Feshbach resonances imposes another restriction on the value of the magnetic field. Knowing the resonances spectrum for corresponding temperatures from work[11] and taking into account depolarization, the authors decided to maintain the magnetic field in the region of 3.65–3.95 G free of resonances.

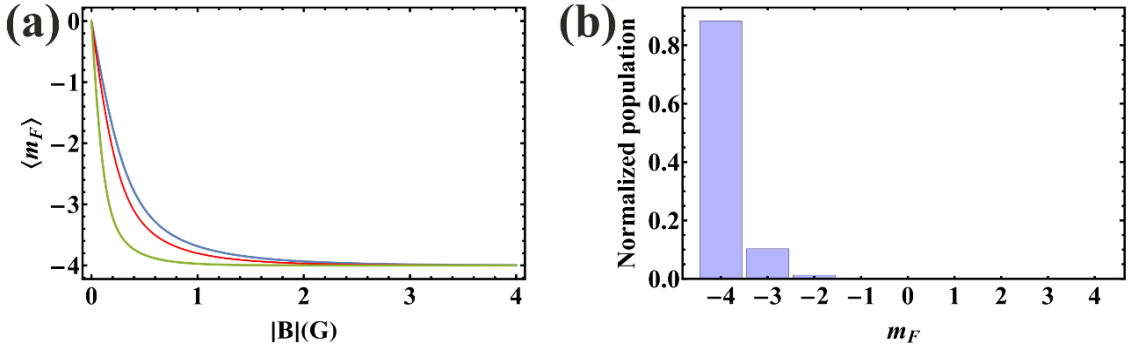

Figure 2. a) Average polarization of the cloud versus holding magnetic field for the green line 20 µK, red line 40 µK, and blue line 50 µK. b) Normalized population of each m_F at holding a magnetic field of 1.2 G with a temperature of 40 µK.

The value of the magnetic field may be maintained by coils, the current of which is synchronized with the shifting the ODT waist. Such an approach requires the knowledge of the magnetic field along the transport trajectory depending on the current in the coils. To measure the magnetic field, the following procedure was performed. The atoms, initially located in the ODT in the main chamber, were transferred to the point $x_{WP}$, where a 200-ms pulse of 530.7 nm resonance laser radiation illuminated the atoms and then they were returned to the main vacuum chamber where the absorption imaging on 410.6 nm transition was performed (see Figure 3a). During the resonance pulse the magnetic field was created by the current in the MC coils. The frequency of the 530.7 nm laser beam was scanned using an acousto-optic modulator (AOM) around the transition frequency. Scanning the frequency of elliptically polarized green light along with subsequent imaging provides the resonance picture where one can determine two transitions $|F=4, m_F=-4\rangle \rightarrow |F=5, m_F=-4\rangle$ (π-transition) and

$|F=4, m_F=-4\rangle \to |F=5, m_F=-3\rangle$ ($\sigma^+$-transition) (see Figure 3b). The spacing between the two resonances is determined by the absolute value of the magnetic field $B$ in the region near $x_{WP}$ according to the relation

$$\nu(|4,-4\rangle \to |5,-3\rangle) - \nu(|4,-4\rangle \to |5,-4\rangle) = \frac{\mu_B g_5}{h} B, \quad (2)$$

where $\nu(|4,-4\rangle \to |5,-3\rangle)$ is the frequency of $\sigma^+$-transition, $\nu(|4,-4\rangle \to |5,-4\rangle)$ – the frequency of $\pi$-transition, $\mu_B$ – the Bohr magneton, $h$ – the Planck constant, $g_5 = 1.008$ – the Lande factor for $F = 5$ level. Figure 3c presents the magnetic field as a function of the current in the MC coils and the ODT waist position. For every $x_{WP}$, the interpolating function is $\sqrt{(kI + B_\parallel)^2 + B_\perp^2}$. Note that the total magnetic field $\sqrt{B_\parallel^2 + B_\perp^2}$ with $I = 0$ was not less than 0.65 G (Figure 3d).

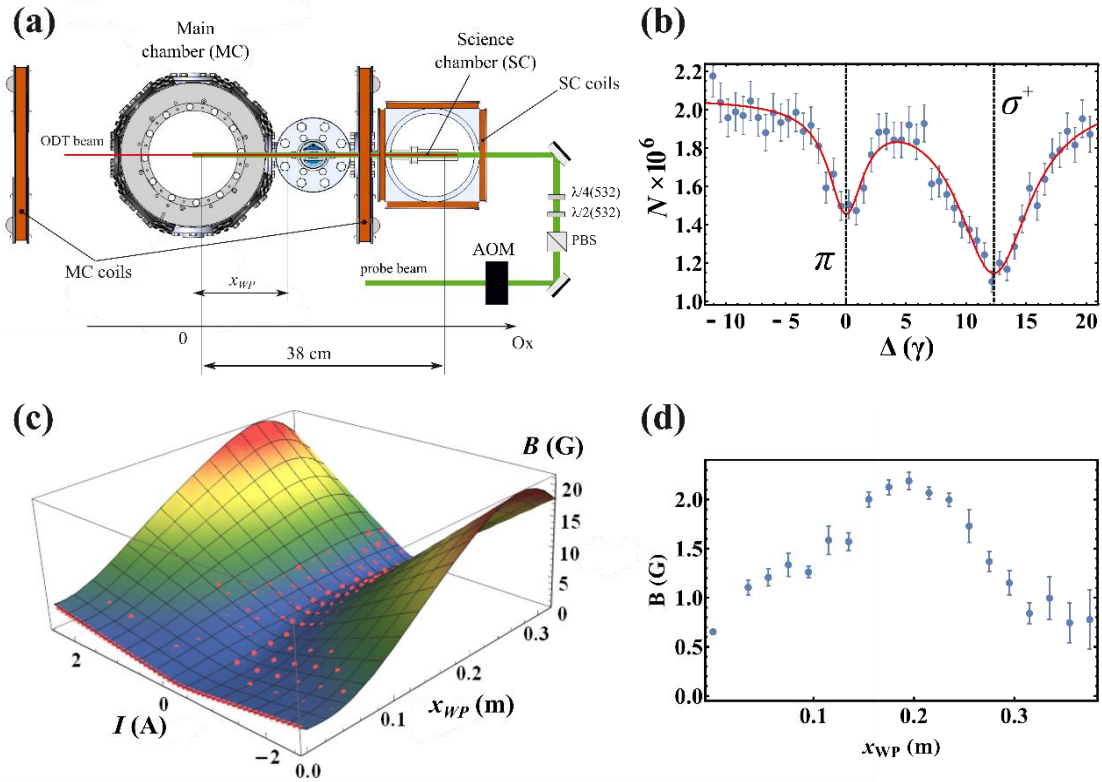

Figure 3. a) Setup of the magnetic field calibration experiment with optical resonance. The Main chamber (MC) coils were used to adjust the magnetic field along the transport trajectory. The science chamber (SC) coils were not used for the transport. The AOM was used to tune the frequency of the probe beam to obtain

the graphs similar to the one in (b). b) Typical optical resonances in the magnetic field calibration experiment. The plot is given in the number of atoms $N$ vs the detuning from atomic resonance $\Delta$ in the units of the natural linewidth of the transition $\gamma$. Solid line is fit of experimental data with sum of two Lorentzian functions. c) Absolute value of the magnetic field as a function of the current in the MC coils and the ODT waist position interpolated by the experimentally obtained points. d) Absolute value of the magnetic field as a function of the ODT waist position $x_{WP}$ when the current in the MC coils $I = 0$ A.

Given the dynamics of the position of the ODT beam waist $x_{WP}(t)$, the dipole force along the $Ox$ axis exerted on an atom with the coordinate $x$ is determined by the potential

$$U(x, x_{WP}(t)) = -\frac{2\alpha P}{\pi \left(w_0(x_{WP}(t))\right)^2 \left[1 + \left(\frac{x_{WP}(t) - x}{x_{REL}}\right)^2\right]}. \qquad (3)$$

Here $w_0(x_{WP}(t))$ is the ODT beam waist depending on its position $x_{WP}(t)$, $\alpha$ – the polarizability of the thulium atom in the ODT, $P$ – the total power of the ODT beam, $x_{REL}$ – the Rayleigh length of the ODT beam. Experimentally, the $x_{WP}(t)$ is set via a focus-tunable lens using a preliminary measured calibration (Figure 1b).

Previously obtained calibration (see Figure 3c) allowed to adjust the spatial profile of the magnetic field along the $Ox$ axis via the current in the MC coils (Figure 3a). The magnetic field exerts a force on magnetic atoms depending on the gradient

$$\vec{F}_m = \nabla(\vec{\mu}\vec{B}) = \nabla(\mu\frac{\vec{B}}{B}\vec{B}) = \mu\nabla B, \qquad (4)$$

where $\mu = m_F \mu_B g_5$ is the magnetic dipole moment of the atom. Note that the force depends only on the gradient of the magnetic field value because the magnetic moments of the atoms are aligned with the field. Consequently, the motion of the center of mass of a cloud of atoms along the $Ox$ axis governed by Newton's second law was considered:

$$m\ddot{x}_{CM}(t) = F_{mx}(x_{CM}(t)) - \nabla U(x, x_{WP}(t))\big|_{x = x_{CM}(t)}, \qquad (5)$$

where $x_{CM}(t)$ – the position of the center of mass of a cloud of atoms along the $Ox$ axis.

The total duration of the transport has several restrictions. A bound from below is set by the characteristic time $T_0 = v_x^{-1}$, where $v_x$ is an ODT frequency along the transport axis: a very fast ramp with time $T \ll T_0$ would be simply ignored by atoms and they would just leave the trap. A bound from above is set by the lifetime of atoms in the ODT measured not to exceed 9 s. Moreover, transferred atoms can oscillate after $x_{WP}$ reaches the final value and stops in time.

The center of mass of a cloud of atoms was set up to move the distance $l$ and duration $T$ with an acceleration changing by $\sin^2(t)$ function[38]. Modeling results and test experiments led to the selection of $T = 20T_0$ to diminish the final oscillations of the atomic cloud. Figure 4a illustrates the chosen functions of the center of mass coordinate $x_{CM}(t)$ and acceleration $a_{CM}(t)$. To maintain the magnetic field at a constant value, that was chosen to be 3.8 G, the measured calibration (Figure 3d) was applied to generate the function $I(t)$ that was set in the MC coils synchronously with the $x_{CM}(t)$ (Figure 4b). By solving equation (5), one can find the necessary profile $x_{WP}(t)$. Figure 4d presents the difference between the $x_{WP}(t)$ and $x_{CM}(t)$ illustrating the fact that the beam waist slightly shifted relative to the actual position of the atoms to exert the optical force. Note that the $F_{mx}$ in the equation (5) depends on the $I(t)$ and is calculated from the calibration function on the Figure 3d. With $x_{WP}(t)$, the optical force would indeed compensate for the magnetic force during the entire transport (see Figure 4c).

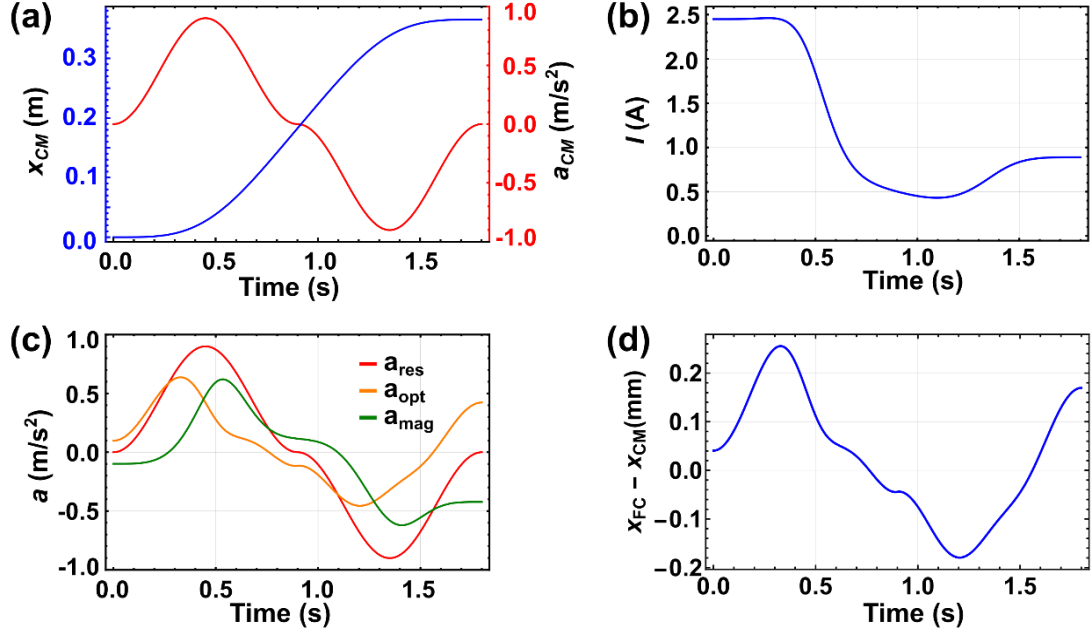

Figure 4. a) The model of the CM motion. The blue line is a coordinate of CM of the cloud versus time, the red line is the acceleration of the CM of the cloud versus time. b) The current in MC coils during the optical transport providing the constant value of the magnetic field (3.8 G) at $x_{CM}(t)$ coordinate at time $t$. c) The acceleration of the CM versus time (red line) and accelerations of CM from magnetic force (green line) and optical dipole force (orange line). d) The difference between $x_{WP}(t)$ and $x_{CM}(t)$.

The obtained $x_{WP}(t)$ function made it possible to move more than $5.4 \cdot 10^6$ atoms to a distance of 38 cm. The efficiency of the transport constituted the value of 0.85 compared to the number of atoms in the ODT measured in the main chamber without transporting them. The decay of atoms after the transport with the magnetic field maintaining reveals only the losses from the background collisions. On the contrary, the decay graph of atoms transferred without maintaining the magnetic field reveals the losses stronger than ones from background collisions (Figure 5). Given the fact that the temperature in the both cases was measured to be approximately the same, the stronger decay without magnetic field correction most likely reveals the depolarization of atoms during the transport [55,56]. Note that even the offset magnetic field exceeding the 0.65 G along the overall transport trajectory (Figure 3d) does not allow to preserve the polarization of atoms. Consequently, the maintaining of the magnetic field significantly increases the initial phase space density for the evaporative cooling. Also, one can try to further improve the efficiency of transport with the machine learning approach[11,57].

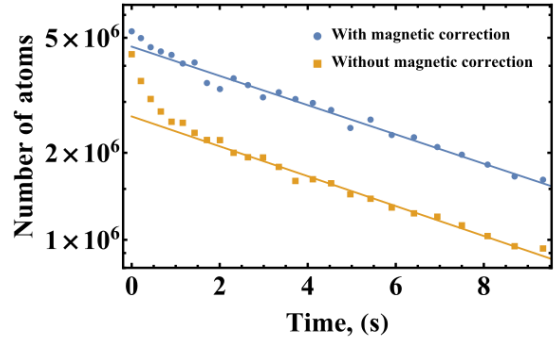

Figure 5. The decay rates of atomic cloud transferred with (blue dots) and without (orange dots) maintaining the magnetic field during transport.

In summary, the optical transport of magnetic thulium atoms was implemented. The measurement of the magnetic field along the transport made it possible to maintain it constant to reduce the losses related to the Fano-Feshbach resonance spectrum and to maintain the $|F=4, m_F=-4\rangle$ state of atoms. The force of the magnetic field gradient was compensated for by a sophisticated ramp for the lens with a tunable focus. As a result, $5.4\cdot 10^6$ atoms were transported to a distance of 38 cm compared to $6.4\cdot 10^6$ atoms remaining in the ODT without transport after the same period of storage time, as the time used for storage.

This work was supported by Rosatom in the framework of the Roadmap for Quantum computing (Contract No. 868-1.3-15/15-2021 dated October 5, 2021).

## AUTHOR DECLARATIONS

### Conflict of Interest

The authors have no conflicts to disclose.

### Author Contributions

**Davlet Kumpilov**: Conceptualization (equal), Data curation (lead), Formal analysis (equal), Investigation (lead), Methodology (lead), Software (supporting), Visualization (equal), Writing – original draft (lead), Writing – review & editing (lead). **Ivan Pyrkh**: Formal analysis (equal), Investigation (equal), Visualization (lead), Writing – original draft (equal). **Ivan Cojocaru**: Software (lead). **Polina Trofimova**: Formal analysis (equal), Investigation (equal), Validation (supporting). **Arjuna Rudnev**: Investigation (equal), Formal analysis (supporting). **Vladimir Khlebnikov**: Validation (equal). **Pavel Aksentsev**: Visualization (supporting). **Ayrat**

**Ibrahimov**: Visualization (supporting). **Alexander Yeremeyev**: Formal analysis (supporting). **Kirill Frolov**: Software (supporting). **Sergey Kuzmin**: Software (supporting). **Anna Zykova**: Resources (equal). **Daniil Pershin**: Methodology (equal), Software (lead). **Vladislav Tsyganok**: Conceptualization (lead), Data curation (lead), Formal analysis (lead), Investigation (lead), Methodology (equal), Project administration (equal), Supervision (equal). **Alexey Akimov**: Conceptualization (equal), Funding acquisition (lead), Project administration (lead), Resources (lead), Supervision (lead), Writing – review & editing (equal).

## DATA AVAILABILITY

The data that support the findings of this study are available from the corresponding author upon reasonable request.